\documentclass[pra,aps,groupaddress,showpacs,prd,preprint]{revtex4}
\usepackage{epsfig,amsmath,graphicx,amssymb}
\def\be{\begin{equation}}
\def\ee{\end{equation}}
\def\bea{\begin{eqnarray}}
\def\eea{\end{eqnarray}}
\def\bes{\begin{subequations}}
\def\ees{\end{subequations}}
\begin{document}
%%%%%%%%%%%%%%%%%%%%%%%%%%%%%%%%%%%%%%%%%%%%%%%%%%%%%%%%%%%%%%%%%%%%%%%%%%%%%%%%%
\title{Linear and Nonlinear Pulse Propagations in Lifetime-Broadened Atomic
Media with Spontaneously Generated Coherence}

\author{Chao Hang}
\affiliation{State Key Laboratory of Precision Spectroscopy and Department of Physics,
East China Normal University, Shanghai 200062, China}
\author{Guoxiang Huang}
\affiliation{State Key Laboratory of Precision Spectroscopy and Department of Physics,
East China Normal University, Shanghai 200062, China}

%%%%%%%%%%%%%%%%%%%%%%%%%%%%%%%%%%%
\date{\today}

%%%%%%%%%%%%%%%%%%%%%%%%%%%%%%%%%%%

\begin{abstract}

The linear and nonlinear pulse propagations in lifetime-broadened
three-state media with spontaneously generated coherence (SGC) are
investigated theoretically. Three generic systems of V-, $\Lambda$-,
and $\Xi$-type level configurations are considered and compared. It
is shown that in linear propagation regime the SGC in the V-type
system can result in a significant change of dispersion and
absorption and may be used to completely eliminate the absorption
and largely reduce the group velocity of a probe field. However, the
SGC has no effect on the dispersion and absorption of the $\Lambda$-
and $\Xi$-type systems. In nonlinear propagation regime, the SGC
displays different influences on Kerr nonlinearity for different
systems. Specifically, it can enhance the Kerr nonlinearity of the
V-type system whereas weaken the Kerr nonlinearity of the
$\Lambda$-type system. Using the SGC, stable optical solitons with
ultraslow propagating velocity and very low pump power can be
produced in the V-type system by exploiting only one laser field in
the system.

\end{abstract}


\maketitle

%%%%%%%%%%%%%%%%%%%%%%%%%%%%%%%%%%%%%%%%%%%%%%%%%%%%%%%%%%%%%%%%%

\section{Introduction}

The propagation of linear and nonlinear light pulses in coherent
atomic media has been an important subject of many recent studies,
such as formation of simultons and adiabatons \cite{KE},
group-velocity reduction \cite{HHDB}, storage and retrieval of
light \cite{LDBH,PFM}, and single-photon pulse propagation
\cite{EAMF}. Among various coherent preparation techniques studied
so far, electromagnetic induced transparency (EIT) is perhaps the
most extensively investigated one because of its diverse practical
applications \cite{FIM}. Due to the quantum interference induced
by a coupling laser field, a probe laser field propagating in an
EIT medium can avoid a large absorption even when it is tuned to a
strong one-photon resonance. EIT media also exhibit drastic change
of dispersion and giant enhancement of Kerr nonlinearity,
resulting in various nonlinear optical phenomena such as
high-efficient frequency conversions \cite{LHL,LKH}, temporal and
spatial optical solitons at very low light level
\cite{WD,MPP,HKH,ZWN}. As a powerful technique, EIT has been used
to study the property of Rydberg atomic ensembles very recently
\cite{Wea}.

In an EIT medium, it is crucial to have at least two laser fields as
they are necessary to be used to create atomic coherence.
Besides the EIT technique, an atomic coherence can also be created by
the quantum interference between two spontaneous emission channels
without using any coupling laser field, which is called the
spontaneously generated coherence (SGC). In recent years, much
attention has been paid to the study on SGC and related topics,
including lasing without inversion \cite{Har,WG,BGS}, coherent
population trapping \cite{men}, spectral narrowing and fluorescence
quenching \cite{ZCL,ZS,PK,KSZ}, fluorescence squeezing \cite{GAC},
giant self-phase modulation \cite{NG}, ground-state quantum beats
\cite{NOB}, cavity-mode entanglement \cite{TLF}, electromagnetically
induced grating \cite{WKJ}, and so on.

Although a large amount of research activities have been made on
SGC, most of them are, however, focused on the static property of
various systems, and only a few works dedicate to the study of pulse
propagation in SGC media. Here we mention the work by Paspalakis
{\it et al.} \cite{PKK} who studied the pulse propagation in a
four-level system, where a ground state is coupled to two closely
spaced excited states by a laser field with both excited states
decaying into a common continuum. Strong, short laser pulses with
adiabaton-like property were observed by using numerical
simulations.

In this article, we investigate, both analytical and numerically,
the linear and nonlinear pulse propagations in lifetime-broadened
three-state media with SGC. In stead of adiabatons, we consider
breather-like nonlinear excitations without using any adiabatic
approximation. Our work includes two aspects: (i) We consider
three generic systems of V-, $\Lambda$-, and $\Xi$-type level
configurations, and compare their linear propagating property with
the SGC effect being taken into account. We show that the SGC in
the V-type system can result in a significant change of dispersion
and absorption and may be used to completely eliminate the
absorption and largely reduce the group velocity of probe field.
However, the SGC has no effect on the dispersion and absorption of
the $\Lambda$- and $\Xi$-type systems. (ii) We demonstrate that in
nonlinear propagation regime, the SGC has different influences on
Kerr nonlinearity for different systems. In particular, it can
enhance the Kerr nonlinearity of the V-type system but weaken the
Kerr nonlinearity of the $\Lambda$-type system. By using the SGC,
stable optical solitons with ultraslow propagating velocity and
very low pump power can be produced in the V-type system. We
stress that the scheme for generating ultraslow optical solitons
presented here is very different from those in Refs.
\cite{WD,MPP,HKH,ZWN} because the suppression of optical
absorption and the reduction of group velocity are not contributed
by an additional control field but by the SGC and hence only a
single laser field is needed.

The work reported here is arranged as follows. In Sec. II,
three-level models of V-, $\Lambda$-, and $\Xi$-type
configurations with SGC are introduced. Linear pulse
propagations are discussed and dispersion and absorption properties
for three different systems are analyzed. In Sec. III, the Kerr
nonlinearity of the V- and $\Lambda$-type systems are analyzed, and
ultraslow optical solitons at very low light level are obtained. In Sec. IV,
a discussion of open system is made. The last section contains a
summary of our main results.

%%%%%%%%%%%%%%%%%%%%%%%%%%%%%%%%%%%%%%%%%%%%%%%%%%%%%%%%%%%%%%%%%
%%%%%%%%%%%%%%%%%%%%%%%%%%%%%%%%%%%%%%%%%%%%%%%%%%%%%%%%%%%%%%%%%
\section{Models and pulse propagation in linear regime}

For comparison and also for completeness, we investigate three generic
three-state systems of V-, $\Lambda$-, and $\Xi$-type level configurations,
which are considered separately in the following.

\subsection{V-type system}

\begin{figure}
\centering
\includegraphics[width=11cm]{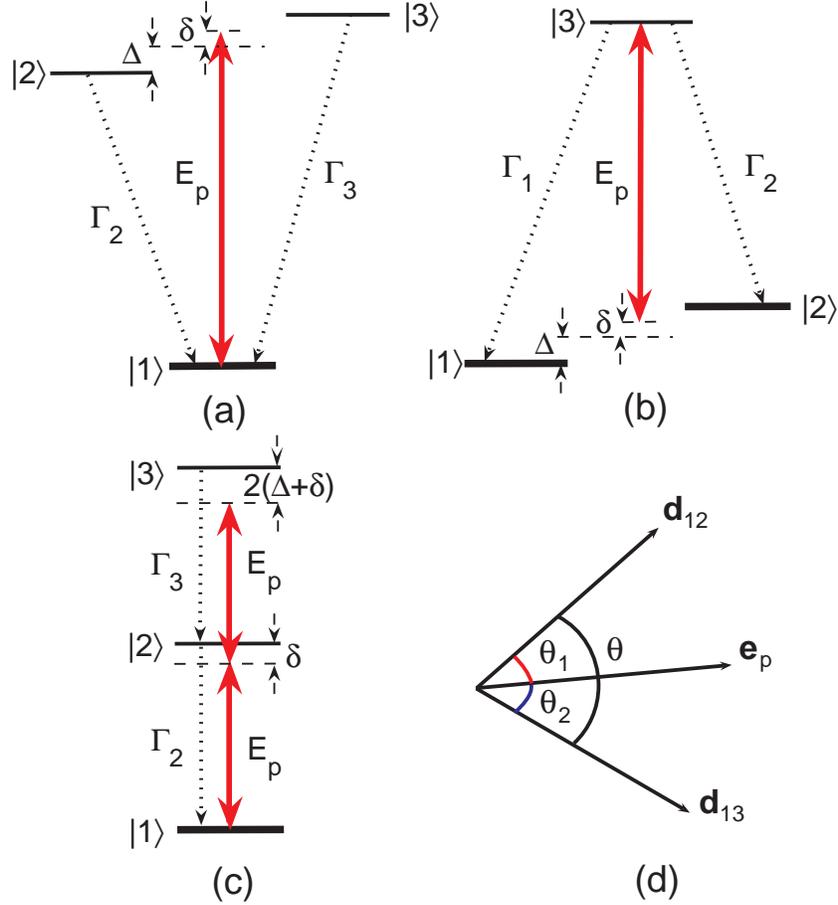}
\caption{(color online) Energy level diagrams and excitation
schemes of lifetime-broadened three-level systems with SGC. (a):
V-type system;  (b): $\Lambda$-type system; (c): $\Xi$-type
system. $|j\rangle$ ($j=1,2,3$) are atomic bare states, ${\bf
E}_p$ is weak probe laser field, $\Delta$ and $\delta$ are
detunings, and $\Gamma_j$ ($j=1,2$) are decay rates of relevant
states. (d): The definition of the alignment angles ($\theta_1$, $\theta_2$) of
the dipole matrix elements (${\bf d}_{12}$, ${\bf d}_{13}$) related to the unit polarization
vector ${\bf e}_p$ of the probe field. $\theta=\theta_1+\theta_2$ is the angle between
${\bf d}_{12}$ and ${\bf d}_{13}$.}
\label{fig1}
\end{figure}

We first consider a three-level V-type atomic system, as shown
in Fig. \ref{fig1}(a), in which two closely spaced excited states $|2\rangle$ and $|3\rangle$
decay simultaneously into the ground state $|1\rangle$ by the spontaneous emission
with decay rates $\Gamma_2$ and $\Gamma_3$, respectively. The quantum interference
between the two decay channels (from $|2\rangle$ to $|1\rangle$ and $|3\rangle$ to $|1\rangle$)
results in the SGC of the system \cite{CRS}.  A weak, pulsed probe field (with duration $\tau_0$) of
center frequency $\omega_{p}$ and wavevector $\textbf{k}_{p}$, i.e.,
\be \label{EF}
{\bf E}_p({\bf r},t)={\bf e}_p {\cal E}_p({\bf r},t) e^{i({\bf k}_p
\cdot {\bf r}-\omega_p t)}+{\rm c.c},
\ee
couples the ground state $|1\rangle$ to the excited states $|2\rangle$ and $|3\rangle$,
where ${\bf e}_p$ and ${\cal E}_p({\bf r},t)$ are the unit polarization vector and envelope function
of the probe field, respectively.
%
%===========================fig1===============================%
%===========================fig1===============================%
%

Under electric-dipole, rotating-wave, and Weisskopf-Wigner approximations,
the equations of motion for the density matrix governing atomic dynamics are
\bes \label{Sch1}
\bea
& & \dot{\rho}_{22}
=-\Gamma_2\rho_{22}+i\Omega_p\rho_{12}-i\Omega_p^{\ast}\rho_{21}
-\eta\frac{\sqrt{\Gamma_2\Gamma_3}}{2}(\rho_{23}+\rho_{32}),\\
& & \dot{\rho}_{33}
=-\Gamma_3\rho_{33}+ip\Omega_p\rho_{13}-ip\Omega_p^{\ast}\rho_{31}
-\eta\frac{\sqrt{\Gamma_2\Gamma_3}}{2}(\rho_{23}+\rho_{32}),\\
& & \dot{\rho}_{21}
=\left[i\left(\Delta+\delta\right)-\frac{\Gamma_2}{2}\right]\rho_{21}
+i\Omega_p(\rho_{11}-\rho_{22})-ip\Omega_p\rho_{23}
-\eta\frac{\sqrt{\Gamma_2\Gamma_3}}{2}\rho_{31},\\
& & \dot{\rho}_{31}
=\left[i\left(-\Delta+\delta\right)-\frac{\Gamma_3}{2}\right]\rho_{31}
+ip\Omega_p(\rho_{11}-\rho_{33})-i\Omega_p\rho_{32}
 -\eta\frac{\sqrt{\Gamma_2\Gamma_3}}{2}\rho_{21},\\
& & \dot{\rho}_{32}
=-\left(i2\Delta+\frac{\Gamma_2+\Gamma_3}{2}\right)\rho_{32}
-i\Omega_p^{\ast}\rho_{31}+ip\Omega_p\rho_{12}
 -\eta\frac{\sqrt{\Gamma_2\Gamma_3}}{2}(\rho_{22}+\rho_{33}),
\eea
\ees
with $\rho_{11}+\rho_{22}+\rho_{33}=1$. Here $\Omega_{p}={\bf
e}_p\cdot {\bf d}_{12}{\cal E}_p/\hbar$ is half Rabi frequency of
the probe field with ${\bf d}_{ij}\equiv \langle i|{\bf
d}|j\rangle$ being the density-matrix elements related to states
$|i\rangle$ and $|j\rangle$, $\Delta=(E_3-E_2)/(2\hbar)$ is half
frequency difference between $|2\rangle$ and $|3\rangle$, and
$\delta=\omega_p-(E_3+E_2)/(2\hbar)$ is one-photon detuning [see
Fig.~\ref{fig1}(a)]. The cross coupling term contributed by the
SGC effect is manifested by the factor
$\eta\sqrt{\Gamma_2\Gamma_3}/2$, with $\eta={\bf d}_{12}\cdot{\bf
d}_{13}/|{\bf d}_{12}||{\bf d}_{13}|=\cos\theta$ denoting the
alignment of two dipole matrix elements ${\bf d}_{12}$ and ${\bf
d}_{13}$, where $\theta$ is the misalignment angle between ${\bf
d}_{12}$ and ${\bf d}_{13}$. If ${\bf d}_{12}$ and ${\bf d}_{13}$
are parallel (i.e. $\theta=0$), one has $\eta=1$, the system
exhibits maximum SGC; if ${\bf d}_{12}$ and ${\bf d}_{13}$ are
perpendicular (i.e. $\theta=\pi/2$), one has $\eta=0$, the system
displays no SGC. $p=|{\bf e}_p\cdot{\bf d}_{13}|/|{\bf
e}_p\cdot{\bf d}_{12}|=|{\bf d}_{13}|\cos\theta_1/[|{\bf d}_{12}|\cos\theta_2]$, where
$\theta_1$ ($\theta_2$) is the misalignment angle between ${\bf d}_{12}$ (${\bf d}_{13}$) and ${\bf e}_p$.
In the following, we assume $|{\bf d}_{13}|\simeq|{\bf
d}_{12}|$ and a particular case can be found that $\theta$ is equally partitioned by
${\bf e}_p$, i.e. $\theta_1 \simeq \theta_2=\theta/2$, as did by Wan et al. \cite{WKJ}. In such case, we have $p\simeq 1$ and
$\eta$ is still kept a free parameter with its value taking between $-1$ and $1$ [see
Fig.~\ref{fig1}(d)].

The equation of motion for the probe-field Rabi frequency
$\Omega_{p}$ can be obtained by using Maxwell equation. Under
slowly-varying envelope approximation, it reads
\be\label{MAX1} i\left(\frac{\partial}{\partial
z}+\frac{1}{c}\frac{\partial}{\partial t}\right)\Omega_{p}+\kappa
(\rho_{21}+p\rho_{31})=0, \ee
where $\kappa={\cal N}_a\omega_{p}|{\bf
e}_p\cdot{\bf d}_{12}|^2/(2\epsilon_0 c \hbar)$ with ${\cal N}_a$
being the atomic concentration. For simplicity, we assume in the
following that $\Gamma_2\approx\Gamma_3\equiv \Gamma$.

The linear optical response of the system can be obtained by solving
the Maxwell-Bloch (MB) Eqs.  (\ref{Sch1}) and (\ref{MAX1}).
Assuming $\Omega_{p}$ is a small quantity,
$\rho_{11}\approx1$, and $\rho_{21}$, $\rho_{31}$, and
$\Omega_p$ are proportional to $\exp[i(Kz-\omega t)]$, we obtain the
linear dispersion relation
\be \label{Disp1}
K(\omega)=\frac{\omega}{c}
-\kappa\left[\frac{\omega+d_3-ip\eta\Gamma/2}{D(\omega)}+
\frac{p^2(\omega+d_2)-ip\eta \Gamma/2}{D(\omega)}\right], \ee
where $D(\omega)=(\omega+d_2)(\omega+d_3)+\eta^2 \Gamma^2/4$ with
$d_2=\Delta+\delta+i\Gamma/2$, and $d_3=-\Delta+\delta+i\Gamma/2$.
By Taylor expanding $K(\omega)$ around $\omega=0$ \cite{note1}, we obtain
$K(\omega)=K_{0}+K_{1}\omega+\frac{1}{2}K_{2}\omega^2+\cdots$, with the
expansion coefficients
$K_{j}=[\partial^{j}K(\omega)/\partial\omega^{j}]|_{\omega=0}$
($j=0$, 1, 2,$\cdots$) explicitly given by
\bes \label{KJ}
\bea
& & \label{K0} K_0=-\kappa\frac{p^2d_2+d_3-ip\eta
\Gamma}{d_2d_3+\eta^2 \Gamma^2/4},
\\
& & \label{K1} K_1=\frac{1}{c}-\kappa\frac{1+p^2}{d_2d_3+\eta^2
\Gamma^2/4}
+\kappa\frac{(p^2d_2+d_3-ip\eta\Gamma)(d_2+d_3)}{(d_2d_3+\eta^2 \Gamma^2/4)^2},\\
& & \label{K2} K_2=2\kappa\frac{(1+p^2)(d_2+d_3)}{(d_2d_3+\eta^2
\Gamma^2/4)^2} +2\kappa\frac{p^2d_2+d_3-ip\eta\Gamma}{(d_2d_3+\eta^2
\Gamma^2/4)^2}
-2\kappa\frac{(p^2d_2+d_3-ip\eta\Gamma)(d_2+d_3)^2}{(d_2d_3+\eta^2
\Gamma^2/4)^3}.
\eea
\ees
Here, $K_{0}={\rm Re}(K_0)+i{\rm Im}(K_0)$ gives the phase shift  per unit
length and absorption coefficient, $K_1$ determines the group velocity $V_g$ ($\equiv 1/K_1$),
and $K_2$ represents the group-velocity dispersion.

\begin{figure}
\centering
\includegraphics[width=12cm]{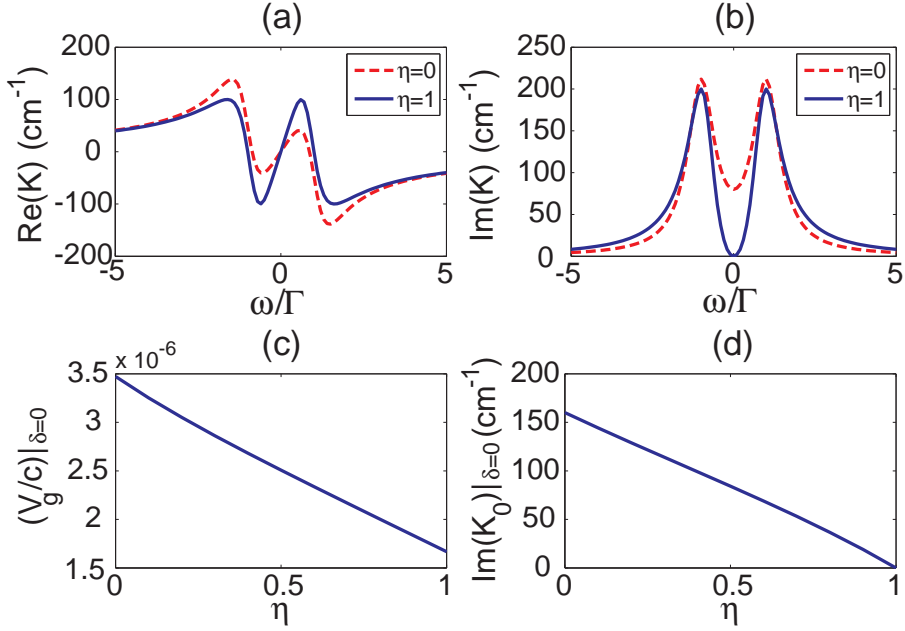}
\caption{(color online) (a), (b): ${\rm Re}(K)$ and ${\rm Im}(K)$
as functions of $\omega/\Gamma$ with maximum SGC (i.e. $\eta=1$;
the solid line) and without SGC (i.e. $\eta=0$; the dashed line).
(c), (d): Im($K_0)|_{\delta=0}$ and $(V_{g}/c)|_{\delta=0}$ as
functions of $\eta$. } \label{fig2}
\end{figure}

Shown in Shown in Fig. \ref{fig2}(a) and Fig. \ref{fig2}(b)
%
%===========================fig2===============================%
%===========================fig2===============================%
%
are respectively the real part and imaginary part of $K$ as
functions of $\omega$,  which characterize the dispersion and
absorption of the system. The solid lines in the figure are for
the case with the maximum SGC (i.e.  $\eta=1$), whereas the dashed
lines are for the case without the SGC (i.e.  $\eta=0$).  System
parameters are chosen as $\kappa=1.0\times10^9$ cm$^{-1}$s$^{-1}$,
$\Gamma=1.0\times10^7$ s$^{-1}$, $\Delta=1.0\times10^7$ s$^{-1}$,
$\delta=0$, and $p=1$. We see that in the region around $\omega=0$
the probe-field displays a drastic change of dispersion (and hence
a drastic reduction of group velocity) (panel (a)\,) and a large
suppression of absorption (panel (b)\,). Obviously,  the reduction
of group-velocity and the suppression of absorption with the SGC
are much more significant than those without the SGC. These can be
seen also by the expression of the group velocity and the
absorption coefficient at $\delta=0$:
\bes
\bea\label{sgc1}
& & V_{g}|_{\delta=0} = \left\{\frac{1}{c}+2\kappa\frac{\Delta^2
                           -(1-\eta)^2\Gamma^2/4}{[\Delta^2+(1-\eta^2)\Gamma^2/4]^2}\right\}^{-1},\\
& & {\rm Im}(K_0)|_{\delta=0} =
\frac{\kappa(1-\eta)\Gamma}{\Delta^2+(1-\eta^2)\Gamma^2/4}, \eea
\ees
which are respectively shown in the panels (c) and (d) of Fig.
\ref{fig2}. Notice that the group velocity of the probe field can
be lowered by increasing the SGC effect (panel (c)\,). If $\eta=1$
one has Im$(K_0)|_{\delta=0}=0$, i.e. the absorption of the probe
field can be completely eliminated by means of the SGC. But if the
SGC is absent (i.e. $\eta=0$), Im$(K_0)|_{\delta=0}$ is not only
non-zero but also with a large positive value, and hence the
probe-field absorption is quite significant (panel (d)\,).
Consequently, in the absence of the SGC the probe filed can not
propagate to a long distance.

It is instructive to discuss the SGC effect on the probe-field
absorption in more details. From Fig. \ref{fig2}(b), we see that
the probe-field absorption is suppressed around $\omega=0$ in both
cases with and without the SGC, resulting in the appearance of
transparency windows in the absorption spectrum. However, the
depth and width of the transparency windows are quite different.
The transparency window with the SGC ($\eta=1$; the solid line) is
much more deeper and wider than that without the SGC ($\eta=0$;
the dashed line).

\begin{figure}
\centering
\includegraphics[width=9cm]{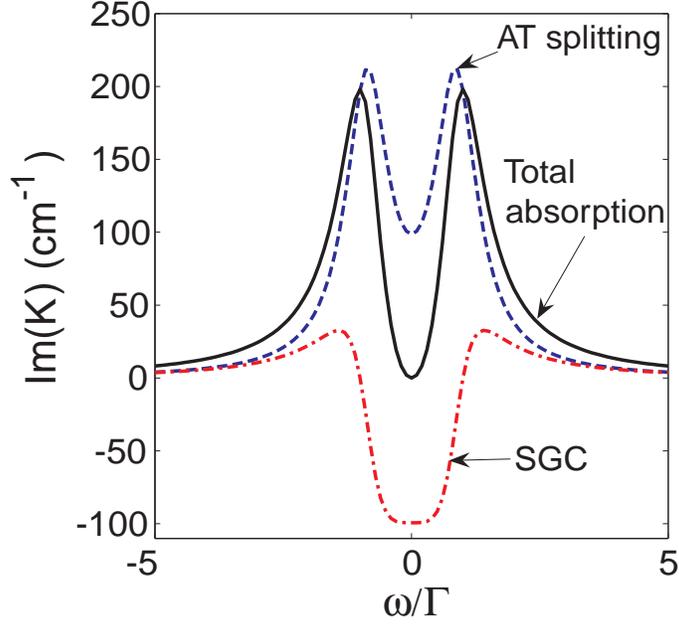}
\caption{(color online) The spectra of AT-splitting (dashed line),
SGC (dotted line), and total absorption (solid line) as functions
of $\omega/\Gamma$ with $\delta=0$.} \label{fig3}
\end{figure}

The difference for the absorption spectra for the above two cases
can be understood more clearly as follows. For illustration, we
split the imaginary part of $K$ into the form
\bea\label{split1} {\rm Im}(K)=&
&\kappa\left[\frac{\Gamma}{2}\left(\frac{1}{(\omega+\delta-v)^2
+\Gamma^2/4}+\frac{1}{(\omega+\delta+v)^2+\Gamma^2/4}\right)\right.\nonumber\\
&&+\left.\frac{\eta\Gamma}{2v}\left(\frac{\omega+\delta-v}{(\omega+\delta-v)^2
+\Gamma^2/4}-\frac{\omega+\delta+v}{(\omega+\delta+v)^2+\Gamma^2/4}\right)\right],
\eea
with $v=\sqrt{\Delta^2-\eta^2\Gamma^2/4}$. The first two terms on
the right hand side (RHS) of Eq. (\ref{split1}) correspond to two
resonances of the excited states $|2\rangle$ and $|3\rangle$,
shown by the dashed line of Fig. \ref{fig3}.
%
%===========================fig3===============================%
%===========================fig3===============================%
%
The dip with a non-zero minimum in the dashed line can be interpreted as a gap between the two resonances,
which is a typical character of Autler-Townes  (AT)  splitting \cite{abi}.
The next two terms (interference terms) provide a gain (absorption) when $\eta>0$ ($\eta<0$),
shown in Fig. \ref{fig3} by the dotted-dashed line. Such terms come from the quantum interference effect
between two spontaneous emission channels (i.e. from $|2\rangle$ to $|1\rangle$ and from $|3\rangle$ to $|1\rangle$),
a typical character of SGC. The sum of the four terms on the RHS of Eq. (\ref{split1}) is the total absorption
of the probe field, shown by the solid line of Fig. \ref{fig3}. From these results we clearly see that
it is the joint contribution from the AT splitting and the SGC effect that make the absorption of the
probe field vanish, and hence providing the possibility for a long-distance transmission of the probe field
in both the linear and nonlinear propagation regimes.

\subsection{$\Lambda$-type system}

We now consider a three-level $\Lambda$-type system, in which a weak probe field with
the form (\ref{EF}) couples the excited state $|3\rangle$ to the two ground states $|1\rangle$ and
$|2\rangle$, with corresponding spontaneous decay rates $\Gamma_1$ and $\Gamma_2$, respectively (see Fig. \ref{fig1}(b)\,).
A SGC occurs by the quantum interference between two spontaneous decay channels
from $|3\rangle$ to $|1\rangle$ and from $|3\rangle$ to $|2\rangle$ \cite{JAV,men1}.

The atomic dynamics of the system is described by the density matrix equations
\bes  \label{Sch-1}
\bea
\dot{\rho}_{11}
=& &\Gamma_1\rho_{33}+ip\Omega_p^{\ast}\rho_{31}-ip\Omega_p\rho_{13},\\
\dot{\rho}_{22}
=& &\Gamma_2\rho_{33}+i\Omega_p^{\ast}\rho_{32}-i\Omega_p\rho_{23},\\
\dot{\rho}_{21}
=& &-i2\Delta\rho_{21}+i\Omega_p^{\ast}\rho_{31}-ip\Omega_p\rho_{23}
+\eta\sqrt{\Gamma_1\Gamma_2}\rho_{33},\\
\dot{\rho}_{31} =&
&\left[-i\left(\Delta+\delta\right)-\frac{\Gamma_1+\Gamma_2}{2}\right]\rho_{31}+ip\Omega_p(\rho_{11}-\rho_{33})
+i\Omega_p\rho_{21},\\
\dot{\rho}_{32} =&
&\left[i(\Delta-\delta)-\frac{\Gamma_1+\Gamma_2}{2}\right]\rho_{32}+i\Omega_p(\rho_{22}-\rho_{33})
+ip\Omega_p\rho_{12}, \eea
\ees
with $\rho_{11}+\rho_{22}+\rho_{33}=1$. Here $\Omega_{p}={\bf
e}_p\cdot {\bf d}_{13}{\cal E}_p/\hbar$ is half Rabi frequency of the
probe field, $\Delta=(E_2-E_1)/(2\hbar)$ is half frequency
difference of two closely spaced ground-state levels, and
$\delta=(2E_3-E_1-E_2)/(2\hbar)-\omega_p$ is one-photon detuning
(see Fig.~\ref{fig1}(b)\,). The SGC effect is still described by
the factor $\eta\sqrt{\Gamma_1\Gamma_2}/2$, with
$\eta={\bf d}_{13}\cdot{\bf d}_{23}/|{\bf d}_{13}||{\bf d}_{23}|$.

The equation of motion for $\Omega_{p}$ is
\be\label{MAX-1} i\left(\frac{\partial}{\partial
z}+\frac{1}{c}\frac{\partial}{\partial t}\right)\Omega_{p}+\kappa
(p\rho_{31}+\rho_{32})=0, \ee
where $p=|{\bf e}_p\cdot{\bf d}_{13}|/|{\bf e}_p\cdot{\bf
d}_{23}|=\cos\theta_1/\cos\theta_2$ ($|{\bf d}_{13}|\simeq|{\bf
d}_{23}|$) and $\kappa={\cal N}_a\omega_{p}|{\bf e}_p\cdot{\bf
d}_{23}|^2/(2\epsilon_0 c \hbar)$. For simplicity, we assume in
the following that $\Gamma_1\approx\Gamma_2=\Gamma$.

From the MB Eqs. (\ref{Sch-1}) and (\ref{MAX-1}) it is easy to obtain the linear dispersion
relation of the system, which reads
\be \label{Disp-1} K(\omega)=\frac{\omega}{c} -\kappa
\left[p^2\frac{\rho_{11}^{(0)}}{\omega-d_1}+\frac{\rho_{22}^{(0)}}{\omega-d_2}\right],
\ee
where $d_1=\Delta+\delta-i\Gamma$, $d_2=-\Delta+\delta-i\Gamma$,
$\rho_{11}^{(0)}=|d_1|^2/(|d_1|^2+p^2|d_2|^2)$, and
$\rho_{22}^{(0)}=p^2|d_2|^2/(|d_1|^2+p^2|d_2|^2)$. As before, we
Taylor expand $K(\omega)$ at $\omega=0$, i.e.
$K(\omega)=K_{0}+K_{1}\omega+\frac{1}{2}K_{2}\omega^2+\cdots$. The
expansion coefficients
$K_{j}=[\partial^{j}K(\omega)/\partial\omega^{j}]|_{\omega=0}$
($j=0$, 1, 2,$\cdots$) are given by
$K_0=\kappa(p^2\rho_{11}^{(0)}/d_1+\rho_{22}^{(0)}/d_2)$,
$K_1=1/c+\kappa(p^2\rho_{11}^{(0)}/d_1^2+\rho_{22}^{(0)}/d_2^2)$,
and $K_2=2\kappa(p^2\rho_{11}^{(0)}/d_1^3+\rho_{22}^{(0)}/d_2^3)$.

Equation (\ref{Disp-1}) consists of two simple Lorentzian terms,
corresponding to the resonances between $|3\rangle$ and
$|1\rangle$ and between $|3\rangle$ and $|2\rangle$, respectively.
Thus only an AT splitting in the absorption spectrum of the probe
field occurs. This tells us in the $\Lambda$-type system the SGC
has no effect on the linear dispersion and absorption spectra of
the system. Shown in
%
%===========================fig4===============================%
\begin{figure}
\centering
\includegraphics[width=12cm]{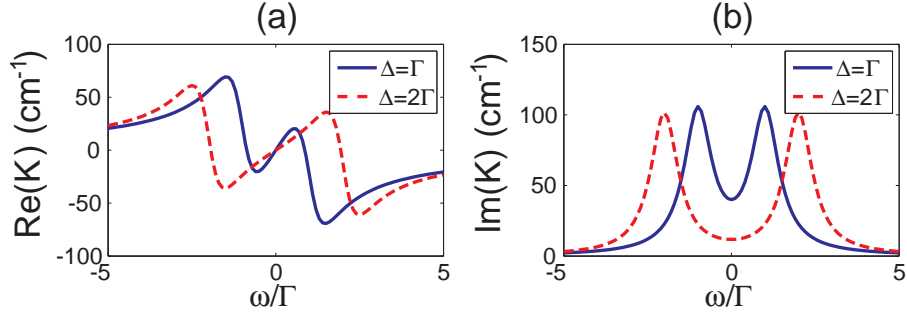}
\caption{(color online) ${\rm Re}(K)$ (panel (a)\,) and ${\rm
Im}(K)$ (panel (b)\,)  as functions of $\omega/\Gamma$ with
$\Delta=\Gamma$ (solid line) and $\Delta=2\Gamma$ (dashed line).}
\label{fig4}
\end{figure}
%===========================fig4===============================%
%
Shown in Fig. \ref{fig4}(a) and Fig. \ref{fig4}(b) are Re($K$) and
Im($K$) as functions of $\omega$. We see that a dip appears in
Im($K$) (panel (b)\,) and, at the same time, the dispersion
changes rapidly in the region around $\omega=0$ ((panel (a)\,). In
addition, Re($K$) and Im($K$) can be adjusted by changing the
value of $\Delta$. The solid (dashed) line in the figure is for
$\Delta=\Gamma$ ($\Delta=2\Gamma$). We see that although Im($K$)
can be lowered by increasing $\Delta$, it has always a large
value. Thus in the present $\Lambda$-type system one can not
acquire a small absorption for the probe field. We must also point
out that the value of $\Delta$ can not be adjusted to be too
large. This is because for a large $\Delta$ the SGC effect
disappears.

\subsection{$\Xi$-type system}

We finally consider a three-level $\Xi$-type system, as shown in Fig. \ref{fig1}(c), in which a weak
probe field couples the ground state $|1\rangle$ to the intermediate state $|2\rangle$ and simultaneously
couples the state $|2\rangle$ to the excited state $|3\rangle$, as suggested in Ref. \cite{FDK}. States $|2\rangle$ and
$|3\rangle$ decay to $|1\rangle$ and $|2\rangle$ with decay rates $\Gamma_1$ and $\Gamma_2$, respectively.
The MB equations governing the evolution of the atoms and the electric field are given by
\bes \label{Sch-3}
\bea
& & \label{sch31} \dot{\rho}_{11}=\Gamma_2\rho_{22}+i\Omega_p^{\ast}\rho_{21}-i\Omega_p\rho_{12},\\
& & \label{sch32} \dot{\rho}_{33}=-\Gamma_3\rho_{33}+ip\Omega_p\rho_{23}-ip\Omega_p^{\ast}\rho_{32},\\
& & \label{sch33} \dot{\rho}_{21}=\left(-i\delta-\frac{\Gamma_2}{2}\right)\rho_{21}
                    +i\Omega_p(\rho_{11}-\rho_{22})+ip\Omega_p^{\ast}\rho_{31}
                    +\eta\sqrt{\Gamma_2\Gamma_3}\rho_{32},\\
& & \label{sch34} \dot{\rho}_{31}=\left[-i2\left(\Delta+\delta\right)-\frac{\Gamma_3}{2}\right]\rho_{31}
                    +ip\Omega_p\rho_{21}-i\Omega_p\rho_{32},\\
& & \label{sch35} \dot{\rho}_{32}=\left[-i(2\Delta+\delta)-\frac{\Gamma_2+\Gamma_3}{2}\right]\rho_{32}
                    +ip\Omega_p(\rho_{22}-\rho_{33})-i\Omega_p^{\ast}\rho_{31},\\
& & \label{sch36} i\left(\frac{\partial}{\partial z}
                    +\frac{1}{c}\frac{\partial}{\partial t}\right)\Omega_{p}+\kappa
                    (\rho_{21}+p\rho_{32})=0,
\eea
\ees
with $\rho_{11}+\rho_{22}+\rho_{33}=1$, where $\Omega_{p}=|{\bf
e}_p\cdot {\bf d}_{12}|{\cal E}_p/\hbar$ is half Rabi frequency of the
probe field, $\Delta=(E_3-2E_2+E_1)/(2\hbar)$ is half frequency
difference of the transitions $|1\rangle\leftrightarrow|2\rangle$
and $|2\rangle\leftrightarrow|3\rangle$, and
$\delta=(E_2-E_1)/\hbar-\omega_p$ is one-photon detuning [see
Fig.~\ref{fig1}(c)]. The last term on the RHS of Eq. (\ref{sch33})
comes from the SGC effect, with
$\eta={\bf d}_{12}\cdot{\bf d}_{23}/|{\bf d}_{12}||{\bf d}_{23}|=\cos\theta$.
In Eq. (\ref{Sch-3}), $p=|{\bf e}_p\cdot{\bf d}_{23}|/|{\bf
e}_p\cdot{\bf d}_{12}|$.

Using the MB Eq. (\ref{Sch-3}) we obtain the linear dispersion relation of the
system
\be \label{Disp0} K(\omega)=\frac{\omega}{c}
-\kappa\frac{1}{\omega-\delta+i\Gamma_2/2}.
\ee
From this formula, we see that: (i) The SGC in the $\Xi$-type system
also does not change the linear dispersion and absorption spectra of the probe
field; (ii) No dip structure in the absorption spectrum is
observed because the system does not possess any AT splitting.
Because the peak of the absorption spectrum locates at the center
frequency of the probe pulse, a long-distance wave propagation
of the probe pulse is not possible in the system. This is not interesting
for wave-propagation problem and hence we discard this model in the
following discussion.

%%%%%%%%%%%%%%%%%%%%%%%%%%%%%%%%%%%%%%%%%%%%%%%%%%%%%%%%%%%%%%%%%%
%%%%%%%%%%%%%%%%%%%%%%%%%%%%%%%%%%%%%%%%%%%%%%%%%%%%%%%%%%%%%%%%%
\section{Pulse propagation in nonlinear regime}

\subsection{V-type system}

Kerr nonlinearity is essential for most nonlinear optical processes.
It can be largely enhanced in resonant optical media,
but usually a serious optical absorption is also accompanied simultaneously.
However, here we show that by the joint action of the AT splitting and the SGC effect,
the Kerr nonlinearity in the V-type system can be enhanced greatly with
the optical absorption eliminated greatly.

The probe-field susceptibility for the V-type system is defined as
\bea
\chi_p&=&\frac{{\cal N}_a|{\bf
e}_p\cdot{\bf d}_{12}|^2}{\epsilon_0
\hbar}\frac{\rho_{21}+p\rho_{31}}{\Omega_p}
\simeq\chi_p^{(1)}+\chi_{p}^{(3)}|{\cal E}_p|^2,
\eea
where $\chi_p^{(1)}$ and $\chi_{p}^{(3)}$ are linear and third-order
susceptibilities, respectively. The real part of $\chi_{p}^{(3)}$
contributes to the Kerr nonlinearity while the
imaginary part of $\chi_{p}^{(3)}$  contributes to the
nonlinear absorption or gain of the system. The explicit expressions
of $\chi_p^{(1)}$ and $\chi_{p}^{(3)}$ can be obtained by solving
Eq. (\ref{Sch1}) under steady-state approximation, which reads
\bes
\label{chi1}
\bea \chi_p^{(1)}&=&-\frac{{\cal N}_a|{\bf
e}_p\cdot{\bf d}_{12}|^2}{\epsilon_0
\hbar}\frac{p^2d_2+d_3-ip\eta\Gamma}{d_2d_3+\eta^2\Gamma^2/4},\\
\chi_{p}^{(3)}&=&\frac{{\cal N}_a|{\bf e}_p\cdot{\bf
d}_{12}|^4}{\epsilon_0 \hbar^3}\frac{d_2A+d_3B-i\eta\Gamma
C/2}{d_2d_3+\eta^2\Gamma^2/4},
\eea
\ees
with the explicit expressions of $A$, $B$, and $C$ being given in
Appendix A.
%
%===========================fig5===============================%
\begin{figure}
\centering
\includegraphics[width=12cm]{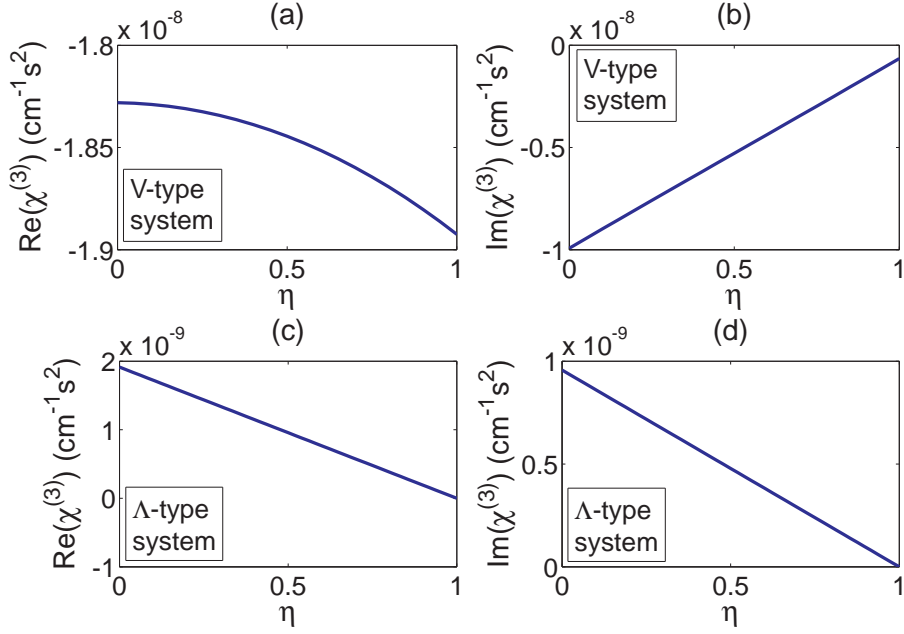}
\caption{(color online) (a) and (b): ${\rm Re}(\chi^{(3)})$ and
${\rm Im}(\chi^{(3)})$ as functions of $\eta$ for  the V-type
system, respectively. (c) and (d): ${\rm Re}(\chi^{(3)})$ and ${\rm
Im}(\chi^{(3)})$ as functions of $\eta$ for  the $\Lambda$-type
system, respectively. } \label{fig5}
\end{figure}
%===========================fig5===============================%
%
Shown in Fig. \ref{fig5}(a) and Fig. \ref{fig5}(b) are ${\rm
Re}(\chi^{(3)})$ and ${\rm Im}(\chi^{(3)})$ as functions of
$\eta$, respectively. System parameters are taken as
${\cal N}_a=10^7$ cm$^{-3}$, $d_{13}\approx
d_{23}=2.5\times10^{-27}$ cm\,C, $\Gamma=1.0\times10^7$ s$^{-1}$,
$\Delta=5.5\times10^7$ s$^{-1}$, $\delta=1.0\times10^7$
s$^{-1}$, and $p=1$. From the figure we see that $|{\rm Re}(\chi^{(3)})|$ grows
as $\eta$ increases, whereas $|{\rm Im}(\chi^{(3)})|$ is reduced as
$\eta$ increases. The condition $|{\rm Re}(\chi^{(3)})|\gg|{\rm
Im}(\chi^{(3)})|$ holds in the whole range of $\eta$. Thus, the SGC
effect enhances the Kerr nonlinearity of the system significantly.
In addition, ${\rm Re}(\chi^{(3)})$ has an order of $10^{-8}$ cm$^{-1}$ s$^2$,
i.e. it is $10^{12}$ times larger than that of conventional
nonlinear optical media \cite{STB}.

The Kerr nonlinearity enhancement obtained above can be used to balance the
dispersion of the system and hence to obtain a lossless and
distortionless optical pulse propagation in nonlinear regime. To this end
we apply the standard multiple-scale method \cite{WD}, which is
beyond the steady-state and adiabatic approximations, to solve Eqs. (\ref{Sch1})
and (\ref{MAX1}). We make the following asymptotic expansion
$\rho_{jj}=\delta_{1j}+\sum^{\infty}_{n=1}\epsilon^{n}\rho_{jj}^{(n)}$
$(j=1,\,2,\,3)$,
$\rho_{ij}=\sum^{\infty}_{n=1}\epsilon^{n}\rho_{ij}^{(n)}$
$(i,j=1,\,2,\,3;\,i\neq j)$, and
$\Omega_{p}=\sum^{\infty}_{n=1}\epsilon^{n}\Omega_{p}^{(n)}$, where
$\epsilon$ is a small parameter characterizing the small population
depletion of the ground state. To obtain a divergence-free
expansion, all quantities on the right hand side of the
expansion are considered as functions of the multi-scale variables
$z_{l}=\epsilon^{l}z$ ($l=0,1,2$) and $t_{l}=\epsilon^{l}t$ ($l=0,
1$). Substituting the expansion and the multi-scale variables into
Eqs. (\ref{Sch1}) and (\ref{MAX1}), we obtain a chain of linear, but
inhomogeneous equations which can be solved order by order.

At the leading order, we get the linear solution
$\Omega_p^{(1)}=F\,\exp \{i[K(\omega)z_{0}-\omega t_{0}]\}$
and the dispersion relation, given by Eq.  (\ref{Disp1}).
At the second order, a divergence-free condition requires
$\partial F /\partial z_{1}+(1/V_{g})\partial
F/\partial t_{1}=0$. Here $F$ is a yet to be determined envelope
function depending on the slow variables $t_{1}$, $z_{1}$ and $z_{2}$.
At the third order, we obtain the nonlinear equation for $F$:
\be \label{NLS1}
i\frac{\partial F}{\partial z_{2}}-\frac{K_2}{2}\frac{\partial^{2}
F}{\partial t_{1}^{2}}-W\exp (-\bar{\beta} z_{2})\,F|F|^{2}=0,
\ee
where $\bar{\beta}=\epsilon^{-2}\beta$ with $\beta=2{\rm Im}(K_0)$
and
\be W=-\kappa\frac{d_2A+d_3B-i\eta\Gamma
C/2}{d_2d_3+\eta^2\Gamma^2/4}.
\ee

After returning to original variables, Eq. (\ref{NLS1}) becomes
\be \label{NLS11}
i\left(\frac{\partial }{\partial
z}+\frac{\beta}{2}\right)U-\frac{K_2}{2}\frac{\partial^{2}
U}{\partial \tau^{2}}-W |U|^{2}U=0,
\ee
where $\tau=t-z/V_{g}$ and $U=\epsilon F e^{-\bar{\beta} z_{2}/2}$.
Equation (\ref{NLS11}) usually has complex coefficients due to
the resonant character of the system. However, as we shall show below,
under the joint action of the SGC and the AT splitting,
practical set of system parameters can be found to make
the imaginary part of the coefficients  be much
smaller than their real part. Then Eq. (\ref{NLS11}) can be approximated
as a nonlinear Schr\"{o}dinger (NLS) equation, which allows soliton
solutions being able to propagate for a rather long distance without
significant attenuation and distortion. The dimensionless form of Eq. (\ref{NLS11})
is
\be\label{NLS2}
i\frac{\partial u}{\partial s}+\frac{\partial^{2} u}{\partial
\sigma^{2}}+2u|u|^{2}=i2\mu u,
\ee
where $s=-z/(2L_D)$, $\sigma=\tau/\tau_0$, and $u=U/U_0$,
$\mu=L_D/L_A$, with $L_{D}=\tau_{0}^{2}/|\tilde{K}_{2}|$ being
the characteristic dispersion length, $L_A=1/\beta$ the characteristic
absorption length, and $U_{0}=(1/\tau_{0})\sqrt{|\tilde{K}_{2}/\tilde{W}|}$ the
characteristic Rabi frequency of the probe field. The symbol tilde
denotes the real part of the corresponding quantity.
%Note that in
%order to find solitons, $U_0$ is obtained by setting $L_D=L_{NL}$,
%i.e., the dispersion and nonlinearity are balanced to each other,
%where $L_{NL}$ is the characteristic nonlinearity length defined by
%$L_{NL}=1/(|\tilde{W}|U_{0}^2)$.

If $L_D$ is much less than $L_A$ (i.e., $\mu\ll1$, which is the case
in the presence of the SGC), the term on the right-hand side of Eq.
(\ref{NLS2}) can be treated as a small perturbation and can be
neglected at the first order. Hence Eq. (\ref{NLS2}) reduces to the
standard NLS equation, which is completely integrable and allows
multi-soliton solutions. After returning to the original variables,
a single soliton solution of the NLS equation corresponds to
\be\label{SOL1}
\Omega_{p}=\frac{1}{\tau_{0}}\sqrt{\frac{\tilde{K}_{2}}{\tilde{W}}}
\,\text{sech}\left[\frac{1}{\tau_0}\left(t-\frac{z}{\tilde{V}_g}\right)\right]\exp\left[i\phi
z-i\frac{z}{2L_D}\right].
\ee

We now present a practical numerical example to support the above
results. Consider a cold alkali atomic gas, for which the system parameters
can be taken as
$d_{24}\approx d_{34}=2.5\times10^{-27}$ cm C,
$\kappa=1.0\times10^9$ cm$^{-1}$ s$^{-1}$, $\Gamma_2\approx
\Gamma_3=1.0\times 10^7$ s$^{-1}$, $\Delta=5.5\times 10^7$ s$^{-1}$,
$\delta=1.0\times10^7$ s$^{-1}$, $\eta=1$, and $p=1$. Then we obtain
$K_0=6.83+i0.23$ cm$^{-1}$, $K_1=(0.73-i0.05)\times 10^{-6}$
cm$^{-1}$s, $K_2=(0.14+i0.06)\times 10^{-13}$ cm$^{-1}$s$^2$, and
$W=(1.00+i0.03)\times 10^{-14}$ cm$^{-1}$s$^2$. Notice that the
imaginary parts of these quantities are indeed much smaller than their
real parts. The reason is that the SGC plays an important role to make
the optical absorption of the system be largely eliminated.
Taking $\tau_0=0.8\times10^{-7}$ s,
we have $L_D=0.45$ cm, $U_0=1.5\times10^7$ $s^{-1}$,
and $L_A=4.3$ cm ($\mu\approx0.1$). Then the propagating velocity of the
soliton is given by
\be
\tilde{V}_g=4.6\times10^{-5}\,c,
\ee
i.e., the soliton formed in the V-type system travels indeed  with
an ultraslow velocity. We must stress that the present scheme for generating
the ultraslow optical soliton needs only one laser field, which is different from
the EIT scheme, where at least two laser fields are required \cite{WD,MPP,HKH,ZWN}.

With the above parameters, it is easy to estimate the peak power of
the ultraslow optical soliton by using the Poynting¡¯s vector, which reads
\be \label{P}
\bar{P}_{\rm peak}=0.65\,\mu W,
\ee
where the cross-section area of the probe laser beam is taken to be
$\pi\times10^{-4}$ cm$^2$. Thus, we see that very low input power is needed for
generating the ultraslow optical soliton due to the enhancement of
Kerr nonlinearity.

\begin{figure}
\centering
\includegraphics[width=12cm]{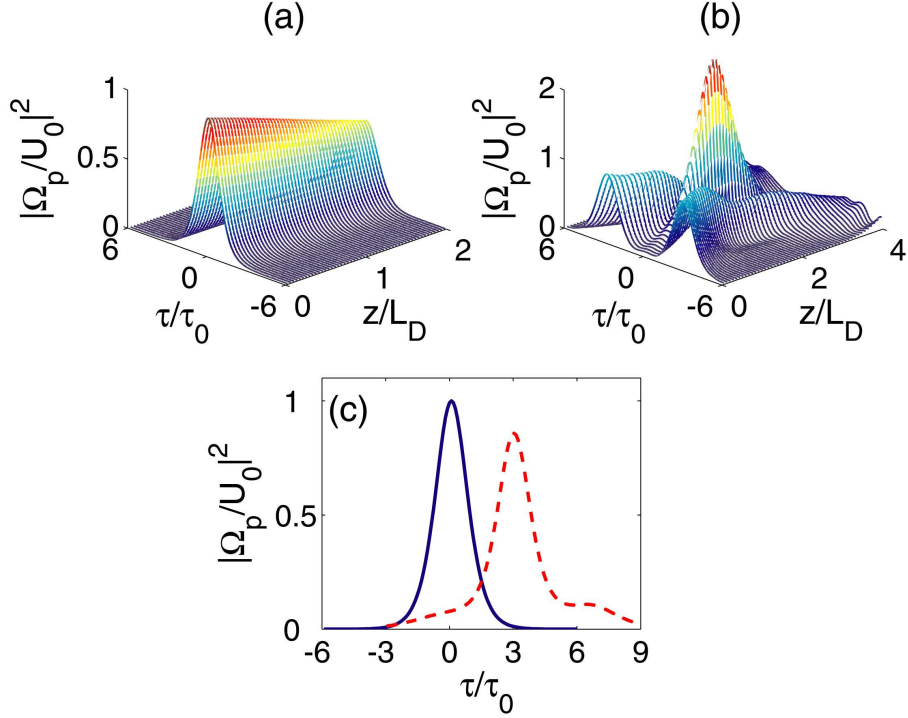}
\caption{(color online) (a): Waveshape of
$|\Omega_p/U_0|^2$ as a function of $z/L_D$ and $\tau/\tau_0$. The
solution is numerically obtained from Eq. (\ref{NLS11}) with full
complex coefficients taken into account. (b): Collision between two
ultraslow solitons. (c): Waveshape of
$|\Omega_p/U_0|^2$ as a function of $\tau/\tau_0$ by directly
integrating Eqs. (\ref{Sch1}) and (\ref{MAX1}) at $z=0$ (solid line)
and $z=2L_D$ (dashed line). } \label{fig6}
\end{figure}

We have also studied the stability of the ultraslow optical soliton
presented above by using numerical simulations. Shown in Fig. \ref{fig6}(a)
%
%===========================fig6===============================%
%===========================fig6===============================%
%
is the wave shape of $|\Omega_p/U_0|^2$ as a function of $z/L_D$
and $\tau/\tau_0$ based on Eq. (\ref{NLS11}).
We see that the amplitude of the soliton undergoes
only a slight decrease and its width undergoes a slight increase after
propagating to a long distance. A simulation of the interaction
between two ultraslow optical solitons is also carried out by initially
inputting two identical solitons [see Fig. \ref{fig6}(b)].
As time goes on, the two solitons collide, pass through, and depart from each
other. They recover basically their initial waveforms after the
collision. Finally, we have also made a numerical simulation by directly
integrating Eqs. (\ref{Sch1}) and (\ref{MAX1}) to confirm the
analytical prediction, with the result shown in Fig. \ref{fig6}(c).

\subsection{$\Lambda$-type system}

The probe-field susceptibility for the $\Lambda$-type system is
\bea
\chi_p=\frac{{\cal N}_a|{\bf
e}_p\cdot{\bf d}_{12}|^2}{\epsilon_0
\hbar}\frac{p\rho_{31}+\rho_{32}}{\Omega_p}
\simeq \chi_p^{(1)}+\chi_{p}^{(3)}|{\cal E}_p|^2,
\eea
where
\label{chi-1}
\bea \chi_p^{(1)}&=&\frac{{\cal N}_a|{\bf
e}_p\cdot{\bf d}_{12}|^2}{\epsilon_0
\hbar}\left(p^2\frac{\rho_{11}^{(0)}}{d_1}+\frac{\rho_{22}^{(0)}}{d_2}\right),\nonumber\\
\chi_{p}^{(3)}&=&\frac{{\cal N}_a|{\bf e}_p\cdot{\bf
d}_{12}|^4}{\epsilon_0
\hbar^3}\left(\frac{-A+B+C}{d_1}+\frac{-2A+B^{\ast}-C}{d_2}\right),
\eea
with $A=ip\rho_{11}^{(0)}(1/d_1^{\ast}-1/d_1)/\Gamma$,
$B=(p\rho_{11}^{(0)}/d_1-p\rho_{22}^{(0)}/d_2^{\ast}-i\eta\Gamma
A/2)/(2\Delta)$, and
$C=[X_1A-p|d_1|^2(d_2B-d_2^{\ast}B^{\ast})-|d_2|^2(d_1^{\ast}B-d_1B^{\ast})]/X_2$
with $X_1=-i\Gamma(2|d_1|^2-p|d_2|^2)$, and
$X_2=i\Gamma(|d_1|^2+p|d_2|^2)$.

Shown in Fig. \ref{fig5}(c) and Fig. \ref{fig5}(d) are ${\rm
Re}(\chi^{(3)})$ and ${\rm Im}(\chi^{(3)})$ as functions of the SGC
parameter $\eta$. The values of the system parameters are the same
with those used for the V-type system. We see that both ${\rm
Re}(\chi^{(3)})$ and ${\rm Im}(\chi^{(3)})$ decreases rapidly as
$\eta$ grows, i.e., the SGC effect weakens the Kerr nonlinearity of
the system. Based on this reason and on the large linear optical
absorption shown in the last section, we conclude that an ultraslow
optical soliton is not possible in the present $\Lambda$-type
system.

%%%%%%%%%%%%%%%%%%%%%%%%%%%%%%%%%%%%%%%%%%%%%%%%%%%%%%%%%%%%%%%%%
\section{The case for open systems}

The results presented above are valid only for close systems.
However, realistic physical systems, in particular molecules \cite{laz},
are usually open ones in which additional spontaneous decay pathways from upper
levels to some other lower levels  exist and hence spoil quantum coherence.
In this section, we examine an open system related to V-type configuration
and show that the additional decay pathways may lead to different initial
population distribution of the states $|1\rangle$, $|2\rangle$, and $|3\rangle$
and hence are detrimental to the SGC of the system.

Consider an extended V-type system
%
%===========================fig7===============================%
\begin{figure}
\centering
\includegraphics[width=9cm]{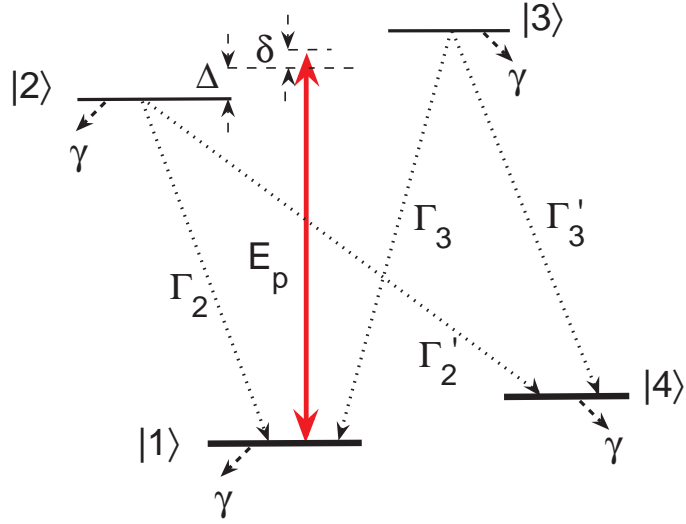}
\caption{(color online) Model scheme of the open system  related
to V-type level configuration. Two
closely spaced upper levels, $|2\rangle$ and $|3\rangle$, decay into
the fourth state $|4\rangle$  with spontaneous decay
rates $\Gamma_2'$ and $\Gamma_3'$, respectively. $|j\rangle$
($j=1,2,3,4$) are bare states, ${\bf E}_p$ is the probe
laser field, $\Delta$ and $\delta$ are detunings.
$\gamma$ is the transient relaxation rate of the particles leaving and
entering light-particle interaction region. } \label{fig7}
\end{figure}
%===========================fig7===============================%
%
with its main part the same as Fig. \ref{fig1}(a), but the two closely
spaced upper levels $|2\rangle$ and $|3\rangle$ may decay
into the fourth state $|4\rangle$ with spontaneous decay rates $\Gamma_2'$ and
$\Gamma_3'$, respectively. In addition, a transient relaxation rate
$\gamma$ is introduced to denote the leaving and entering of particles
in light-particle interaction region; see Fig. \ref{fig7}.
Notice that in this simplified description the state $|4\rangle$
represents many other ground-state sublevels that provide various relaxation
pathways from the two upper sates $|2\rangle$ and $|3\rangle$.
The equations of motion for the density matrix of the system are
\bes \label{Sch2}
\bea
& & \label{Sch2a}
\dot{\rho}_{11}=-\gamma(\rho_{11}-\rho_{11}^{\rm eq})+\Gamma_2\rho_{22}+\Gamma_3\rho_{33}-i\Omega_p\rho_{12}+i\Omega_p^{\ast}\rho_{21}
-ip\Omega_p\rho_{13}+ip\Omega_p^{\ast}\rho_{31}\nonumber\\
& & \hspace{1.1cm}+\eta\sqrt{\Gamma_2\Gamma_3}(\rho_{23}+\rho_{32}),\\
& & \label{Sch2b} \dot{\rho}_{22}
=-\gamma(\rho_{22}-\rho_{22}^{\rm eq})-(\Gamma_2+\Gamma_2')\rho_{22}+i\Omega_p\rho_{12}-i\Omega_p^{\ast}\rho_{21}
-\eta\frac{\sqrt{\Gamma_2\Gamma_3}}{2}(\rho_{23}+\rho_{32}),\\
& & \label{Sch2c} \dot{\rho}_{33}
=-\gamma(\rho_{33}-\rho_{33}^{\rm eq})-(\Gamma_3+\Gamma_3')\rho_{33}+ip\Omega_p\rho_{13}-ip\Omega_p^{\ast}\rho_{31}
-\eta\frac{\sqrt{\Gamma_2\Gamma_3}}{2}(\rho_{23}+\rho_{32}),\\
& & \label{Sch2d}
\dot{\rho}_{44}=-\gamma(\rho_{44}-\rho_{44}^{\rm eq})+\Gamma_2'\rho_{22}+\Gamma_3'\rho_{33},\\
& & \label{Sch2e} \dot{\rho}_{21}
=\left[i\left(\Delta+\delta\right)-\frac{\Gamma_2+\Gamma_2'}{2}\right]\rho_{21}
+i\Omega_p(\rho_{11}-\rho_{22})-ip\Omega_p\rho_{23}
-\eta\frac{\sqrt{\Gamma_2\Gamma_3}}{2}\rho_{31},\\
& & \label{Sch2f} \dot{\rho}_{31}
=\left[i\left(-\Delta+\delta\right)-\frac{\Gamma_3+\Gamma_3'}{2}\right]\rho_{31}
+ip\Omega_p(\rho_{11}-\rho_{33})-i\Omega_p\rho_{32}
 -\eta\frac{\sqrt{\Gamma_2\Gamma_3}}{2}\rho_{21},\\
& & \label{Sch2g} \dot{\rho}_{32}
=-\left(i2\Delta+\frac{\Gamma_2+\Gamma_2'+\Gamma_3+\Gamma_3'}{2}\right)\rho_{32}
-i\Omega_p^{\ast}\rho_{31}+ip\Omega_p\rho_{12}\nonumber\\
& & \hspace{1.1cm} -\eta\frac{\sqrt{\Gamma_2\Gamma_3}}{2}(\rho_{22}+\rho_{33}),
\eea
\ees
with $\rho_{4j}=0$ ($j=1,2,3$). Here $\rho_{jj}^{\rm eq}$ ($j=1,2,3,4$) is the
population of thermal equilibrium when the light field and the spontaneous emission
are absent. The equation of motion
for the probe-field Rabi frequency $\Omega_p$ is still given by Eq.
(\ref{MAX1}). Note that $\sum_{j=1}^3\rho_{jj}\neq 1$ is broken for the present system.

When the probe field is absent, the steady-state solution of the system reads
\bes\label{pop}
\bea
& &
\rho_{11}=\frac{\Gamma(\rho_{22}+\rho_{33})+\eta\Gamma(\rho_{32}+\rho_{23})}{\gamma}+\rho_{11}^{\rm eq},\\
& & \rho_{22}=\frac{M_1\rho_{22}^{eq}+M_2\rho_{33}^{\rm eq}}{M},\\
& & \rho_{33}=\frac{M_1\rho_{33}^{eq}+M_2\rho_{22}^{\rm eq}}{M},\\
& & \rho_{32}=\frac{(\gamma+\Gamma+\Gamma')(2i\Delta-\Gamma-\Gamma')\eta\Gamma/2}{M}\gamma(\rho_{22}^{eq}+\rho_{33}^{\rm eq}),\\
& &
\rho_{44}=\frac{\Gamma'(\rho_{22}+\rho_{33})}{\gamma}+\rho_{44}^{\rm eq},
\eea
\ees
with other $\rho_{jl}=0$, where
$M_1=4\gamma(\gamma+\Gamma+\Gamma')\Delta^2+\gamma(\Gamma+\Gamma')
[(\Gamma+\Gamma')^2+\gamma(\Gamma+\Gamma')-\eta^2\Gamma^2/2]$,
$M_2=\gamma(\Gamma+\Gamma')\eta^2\Gamma^2/2$, and
$M=-(\Gamma+\Gamma')(\gamma+\Gamma+\Gamma')\eta^2\Gamma^2
+(\gamma+\Gamma+\Gamma')^2[(\Gamma+\Gamma')^2+4\Delta^2]$.
For simplicity we have assumed
$\Gamma_2\approx\Gamma_3=\Gamma$ and $\Gamma_2'\approx\Gamma_3'=\Gamma'$.
One sees that there are non-zero initial population in the states $|2\rangle$,
$|3\rangle$, and $|4\rangle$ due to non-zero $\Gamma_2^{\prime}$,
$\Gamma_3^{\prime}$, and $\rho_{jj}^{\rm eq}$.
%Because the energy of the states $|2\rangle$ and $|3\rangle$ is much higher than that of the states
%$|1\rangle$ and $|4\rangle$, one can take $\rho_{22}^{\rm eq}=\rho_{33}^{\rm eq}\approx
%0$, and hence one has $\rho_{11}=\rho_{11}^{\rm eq}$, $\rho_{44}=\rho_{44}^{\rm eq}$,
% and $\rho_{22}=\rho_{33}=\rho_{32}\approx 0$.

It is easy to obtain the linear dispersion relation of the system
\bea \label{Disp2}
K(\omega)&=&\frac{\omega}{c}
-\kappa\left[\frac{(\omega+g_3)(\rho_{11}-\rho_{22})
-ip\eta\Gamma(\rho_{11}-\rho_{33})/2}{D(\omega)}\right.\nonumber\\
& & \left.+ \frac{p^2(\omega+g_2)(\rho_{11}-\rho_{33})-ip\eta
\Gamma(\rho_{11}-\rho_{22})/2}{D(\omega)}\right], \eea
where $D(\omega)=(\omega+d_2)(\omega+d_3)+\eta^2 \Gamma^2/4$ with
$d_2=\Delta+\delta+i(\Gamma+\Gamma')/2$ and
$d_3=-\Delta+\delta+i(\Gamma+\Gamma')/2$.
%
%
%===========================fig8===============================%
\begin{figure}
\centering
\includegraphics[width=12cm]{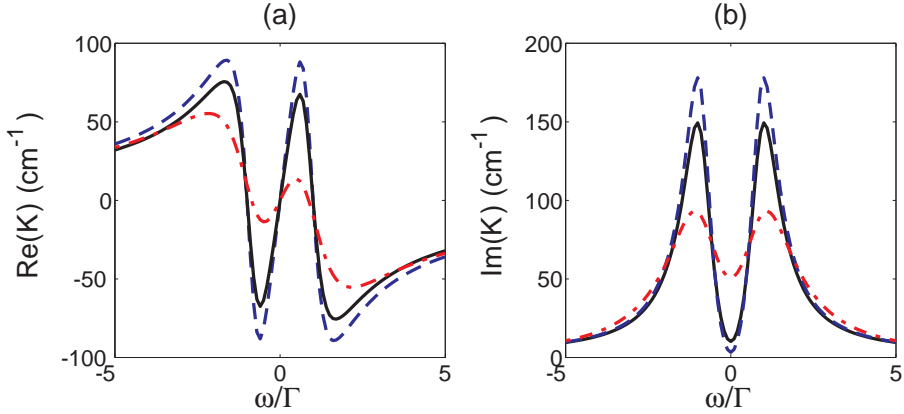}
\caption{(color online) ${\rm Re}(K)$ (panel (a)\,) and ${\rm Im}(K)$ (panel (b)\,)
as functions of $\omega/\Gamma$ for $\Gamma'=0.01\Gamma$ (dashed
line), $\Gamma'=0.1\Gamma$ (solid line), and $\Gamma'=\Gamma$
(dash-dotted line). } \label{fig8}
\end{figure}
%===========================fig8===============================%
%
Illustrated in Fig. \ref{fig8}(a) and Fig. \ref{fig8}(b) are
respectively the spectra of Re($K$)  and Im($K$)  as functions of
$\omega$  for $\Gamma'=0.01\Gamma$ (the dashed line),
$\Gamma'=0.1\Gamma$ (the solid line), and $\Gamma'=\Gamma$ (the
dash-dotted line). When plotting the figure, the value of $\gamma$
is chosen to be $ 10^{-2}\Gamma$ \cite{ZWG}. In addition, $\eta$
is taken to unity in order to show clearly the influence of the
decay rates $\Gamma_2'$ and $\Gamma_3'$ with the maximum SGC. From
Eq. (\ref{Disp2}) and Fig. \ref{fig8} we obtain the following
conclusions: (i) Due to the open character of the system (i.e. the
existence of the additional spontaneous decay pathways from the
states $|2\rangle$, $|3\rangle$ to the state $|4\rangle$), the
slope of Re($K$) curve with respect to $\omega$ decreases when
$\Gamma'$ increases, which means that the group velocity of the
probe field becomes larger when $\Gamma'$ grows; see Fig
\ref{fig8}(a). (ii) The absorption of the probe field can not be
completely eliminated by means of the SGC effect, i.e. the system
is always absorptive. In particular, the minimum of the absorption
spectrum increases when $\Gamma'$ increases; see Fig
\ref{fig8}(b). This tells us that the additional spontaneous decay
pathways in the open system are detrimental to the SGC. Such
conclusions can also be understood in another way. In order to
suppress the absorption of the probe field, it is necessary to
keep the most of atoms populated in the ground state $|1\rangle$.
Nevertheless, from Eq. (\ref{pop}) we see that the additional
spontaneous decay pathways distribute the population of the ground
state into other ground-state sublevels. In addition, when
$\Gamma'$ grows, the population in the state $|4\rangle$ also
grows, which results in the destruction of the quantum coherence
of the system. Consequently, a quantum coherence with zero
(linear) absorption based on the SGC takes place only in closed
systems. However, as shown in Fig. \ref{fig8}(b), although in an
open system a high-quality quantum coherence can not be realized,
a significant suppression of probe-field absorption can still
occur, as shown by the recent experimental observation in a
molecular system \cite{laz}.

%%%%%%%%%%%%%%%%%%%%%%%%%%%%%%%%%%%%%%%%%%%%%%%%%%%%%%%%%%%%%%%%%
\section{Discussion and summary}

Notice that the SGC occurs in systems having near-degenerated
levels with the same angular momentum quantum numbers $J$ and
$m_J$ and nonorthogonal dipole moments, which are rarely satisfied
for realistic atomic systems \cite{XYZ}. However, such type of
quantum interference can be observed in many other systems such as
semiconductor quantum wells and quantum dots \cite{FCS,SCGI,WGX},
autoionizing media \cite{Nak}, and anisotropic vacuum \cite{Aga}.
Our theoretical approach presented above can be easily generalized
to these systems with the SGC.

In summary, in this work we have investigated the linear and
nonlinear pulse propagations in lifetime-broadened three-state media
with SGC. Three generic systems of V-, $\Lambda$-, and $\Xi$-type
level configurations have been considered and compared. We have
shown that in the linear propagation regime the SGC in the V-type system
can result in a significant change of dispersion and absorption and
may be used to completely eliminate the absorption and largely
reduce the group velocity of the probe field. However, the SGC has no
effect on the dispersion and absorption of the $\Lambda$- and
$\Xi$-type systems. We have also shown that in the nonlinear propagation
regime, the SGC displays different influences on Kerr nonlinearity
for different systems. In particular, it can enhance the Kerr
nonlinearity of the V-type system whereas weaken the Kerr
nonlinearity of the $\Lambda$-type system. By exploiting the SGC effect,
stable optical solitons with ultraslow propagating velocity and very
low pump power can be produced in the V-type system by using only a single
laser field in the system.

%%%%%%%%%%%%%

\acknowledgments

This work was supported by the National Natural Science Foundation of China
under Grant No. 10874043, and by the Open Fund from the State Key Laboratory
of Precision Spectroscopy, ECNU.

\appendix

\section{The expressions of $A$, $B$, and $C$ in $\chi_p^{(3)}$}

The expressions of $A$, $B$, and $C$ in $\chi_p^{(3)}$ and $W$ are
given as $A=pX_1+2pX_2+X_3^{\ast}$, $B=2X_1+X_2+pX_3$, and
$C=(2+p)X_1+(2p+1)X_2+pX_3+X_3^{\ast}=A+B$, where
\begin{widetext}
\bea
& &
X_1=\frac{2(Y_1-Y_2)\zeta^2+[\Gamma(Y_3+Y_4)+i2\Delta(Y_3-Y_4)]\zeta
-(\Gamma^2+4\Delta^2)Y_1}{\Gamma(\Gamma^2+4\Delta^2-4\zeta^2)},\nonumber\\
& &
X_2=\frac{2(Y_2-Y_1)\zeta^2+[\Gamma(Y_3+Y_4)+i2\Delta(Y_3-Y_4)]\zeta
-(\Gamma^2+4\Delta^2)Y_2}{\Gamma(\Gamma^2+4\Delta^2-4\zeta^2)},\nonumber\\
& & X_3=\frac{2(Y_3-Y_4)\zeta^2+[(\Gamma+i2\Delta)(Y_1+Y_2)]\zeta
-\Gamma(\Gamma+2i\Delta)Y_3}{\Gamma(\Gamma^2+4\Delta^2-4\zeta^2)},
\eea \end{widetext}
with
\bea
& & Y_1=-\frac{d_3-ip\zeta}{d_2d_3+\zeta^2}
+\frac{d_3^{\ast}+ip\zeta}{d_2^{\ast}d_3^{\ast}+\zeta^2},\nonumber\\
& & Y_2=-\frac{p^2d_2-ip\zeta}{d_2d_3
+\zeta^2}+\frac{p^2d_2^{\ast}+ip\zeta}{d_2^{\ast}d_3^{\ast}+\zeta^2},\nonumber\\
& & Y_3=-\frac{d_3-ip\zeta}{d_2d_3
+\zeta^2}+\frac{p^2d_2^{\ast}+ip\zeta}{d_2^{\ast}d_3^{\ast}+\zeta^2},\nonumber\\
& & Y_4=-\frac{p^2d_2-ip\zeta}{d_2d_3
+\zeta^2}+\frac{d_3^{\ast}+ip\zeta}{d_2^{\ast}d_3^{\ast}+\zeta^2},
\eea
and $\zeta=\eta\Gamma/2$.

%%%%%%%%%%%%%%%%%%%%%%%%%%%%%%%%%%%%%%%%%%%%%%%%%%%%%%%%%%%%%%

%%%%%%%%%%%%%%%%%%%%%%%%%%%%%%%%%%%%%%%%%%%%%%%%%%%%%%%%%%%%%%


\begin{references}
\bibitem{note1}The frequency and wavenumber of the probe field are given by
$\omega_p+\omega$ and $k_p+K(\omega)$, respectively. Thus $\omega=0$ corresponds to
the center frequency of the probe field.
\end{references}
\end{document}